\documentstyle[aps,preprint,epsf,floats]{revtex}
\tighten
\begin{document}
\title{A geometrical origin for the covariant entropy bound}
\author{H. Casini}
\address{{\sl Centre de Physique Th\'{e}orique}, Campus de Luminy, F-13288,
Marseille, France \\
e-mail: casini@cpt.univ-mrs.fr}
\maketitle

\begin{abstract}
Causal diamond-shaped subsets of space-time are naturally 
associated with operator algebras
in quantum field theory, and they are also related to the Bousso
covariant entropy bound. In this work we argue that the net of these causal
sets to which are assigned the local operator algebras of quantum theories
should be taken to be non orthomodular if there is some lowest scale for the
description of space-time as a manifold. This geometry can be related
to a reduction in the degrees of freedom of the holographic type under
certain natural conditions for the local algebras. A non orthomodular net of
causal
sets that implements the cutoff in a covariant manner is constructed. It
gives an explanation, in a simple example, of the non positive expansion
condition for light-sheet selection in the covariant entropy bound. It also
suggests a different covariant formulation of entropy bound.
\end{abstract}


\section{Introduction}

The discovery that black holes have an associated entropy given by the
Bekenstein formula $S=A/(4G\hbar)$ in terms of the horizon area $A$ is
arguably a very important clue to the understanding of the role of gravity
at the quantum level. Current explanations of the black hole entropy vary
between the use of definite models for quantum gravity, and effective ideas
as entanglement entropy and induced gravity (for a general review and
references see \cite{wald}).

The fact that the black hole entropy increases proportionally to the area in
Planck units rather than the volume, has led to the idea of the holographic
or spherical entropy bound \cite{st,bousso1}. This states that the entropy
for a system enclosed in a given approximately spherical surface of area $A$
is less than $A/(4G\hbar)$. The spherical bound can be seen as a statement
on the metastability of macroscopic black holes, that is, that no system
enclosed inside a spherical area $A$ can have greater entropy than a black
hole of the same area \cite{wald}. However, the bound does not work for
irregular surfaces, surfaces inside black holes or in cosmological
situations, where there is a strong time dependence of the matter system
compared with its typical radius (see for details \cite{wald,bousso1} and
references there in).

An appropriate generalized version of this bound to cosmological situations
and general space-times was developed by Bousso in \cite{bousso}, where it 
was called
the covariant entropy bound. Here we briefly introduce it, for a detailed
account see \cite{bousso1}.

Given a spatial codimension two surface $\Omega$ it is possible to construct
four congruences of null geodesics orthogonal to $\Omega$, two past and two
future directed. Suppose that one of these null congruences orthogonal to $
\Omega$ has non positive expansion $\theta$ at $\Omega$. Then call $H$ the
subset of the hypersurface generated by the congruence where the expansion
 is non positive. The hypersurface $H$ is called a light-sheet of $\Omega$.
Figure 1(a) shows an example. The covariant entropy bound
states that the entropy in $H$ is less than $A(\Omega)/(4G\hbar)$.

The covariant bound should be regarded as tentative. However, there are no
known reasonable counterexamples. Indeed, the bound can be shown to be true
in the classical regime under certain conditions equivalent to a local
cutoff in energy, and when the metric satisfies the Einstein equations 
\cite{fmw}. This includes a vast set of physical situations.

Under the conditions of the covariant entropy bound, the naive picture
coming from hyperbolic equations of motion and the Cauchy surfaces in
space-time results drastically changed. In the usual picture the dataset on
the Cauchy surface is to be taken as arbitrary, giving place to an
independent degree of freedom for Planck volume and a maximum number of
states of the order of the exponential of the volume in the cutoff units.
The existence of covariant laws of evolution imposes that the physics inside
the whole causal development of the surface $C$ in Fig. 1(a) should be
described in terms of
the same degrees of freedom. However, the entropy bound
imply that the maximum number of states is further reduced to be some
exponential of the area of the surface $\Omega $. The idea that the physics
inside a given volume (and then in the whole diamond shaped region $S$ of
Fig. 1(a)) would admit a description in terms of independent degrees of
freedom at the bounding surface is known as the holographic principle 
\cite{st}. This reduction would not apply in this simple form when the 
surface $\Omega $ 
is trapped or antitrapped, that is, when the two null congruences
orthogonal to $
\Omega $ having negative expansion are both future or past directed (see
fig. 1(b)). There, the bound is saved by the choosing of negative expansion
light-sheets and the formation of a singularity in space-time.

\begin{figure}
\centering
\leavevmode
\epsfysize=5cm
\bigskip
\epsfbox{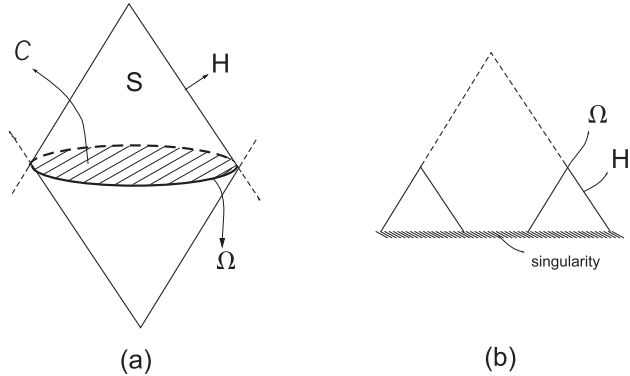}
\caption{(a) The future directed light-sheet $H$ generated by a spatial one
dimensional closed surface $\Omega$ in three dimensional Minkowski space. The
past and future directed light-sheets bound a diamond shaped set $S$ that is
causally closed. This set coincides with the causal development of
a two dimensional spatial surface $C$ that is a Cauchy surface for $S$
(shaded
surface). (b) Flat cosmological model with initial singularity.
The figure shows a cut in the $(t,x)$ plane. A closed spatial
 surface $\Omega$ is represented by two points. The case shown is
when this surface is bigger than the horizon, then it is antitrapped. The
past directed orthogonal null congruences are light-sheets, and will crush
into the singularity. The future directed ones are not light-sheets since
they are initially expanding (i.e. the surface shown with the dashed line). }
\end{figure}

In a quantum field theory the Reeh-Schlieder theorem \cite{rs} impedes the
discussion of the physics in a finite region in terms of subspaces of whole
the Hilbert space, because the fields restricted to a bounded region
generate the whole Hilbert space when acting on the vacuum~\footnote{
I thank C. Rovelli for pointing me this theorem}. However, local algebras of
operators can be defined. The diamond shaped sets of Fig. 1(a) play an
important role in the algebraic approach to quantum field theory
\cite{he,ext}.
In this context, to some sets $S$ in space-time it is associated an algebra $
{\cal A}(S)$ of bounded operators acting on the Hilbert space ${\cal H}$,
(in fact a von Neumann algebra). These have to be regarded as the algebras
generated by the quantum fields averaged using weight functions with support
inside
the given region (see \cite{he} for the relation with conventional quantum
field theory). The operator algebras are local,
in the sense that given two sets $S$,
 and $T$
\begin{equation}
S\subseteq T\Rightarrow {\cal A}(S)\subseteq {\cal A}(T)\ .  \label{a}
\end{equation}
In addition, the operators on the algebra corresponding to the causal
complement $S^{\prime}$ of $S$, ${\cal A}(S^{\prime })$, commute with
the operators in ${\cal A}(S)$. Here the causal complement or opposite 
$S^{\prime}$
of $S$ is the set of points spatially separated from $S$. Then
\begin{equation} {\cal A}(S^{\prime })\subseteq {\cal A}(S)^{\prime }\ , 
\label{b} \end{equation}
 where ${\cal A}^{\prime }$ is the algebra of the
operators that commute with all the operators in ${\cal A}$.

The conditions (\ref{a}) and (\ref{b}) are minimal for the net of algebras,
and some evolution law has to be supplemented. In Minkowski space the
evolution is dictated by the existence of a unitary representation of the
Poincare group acting on the local algebras. 
For a more general situation
the dynamical law would be manifest in that for a set
$S$ \cite{he,ext}
\begin{equation}
 {\cal A}(S)={\cal
A}(S^{\prime\prime})\ .  \label{c}
 \end{equation} 
Then, a natural definition of the diamond shaped sets in
this context, is the sets that satisfy 
$S=S^{\prime \prime}$.
These are called causally closed sets. For example, the diamond shaped set
$S$ in Fig. 1(a) is the domain of dependence $D(C)$ of the surface $C$, that
is, the maximal set where it is possible to determine the variables for a
wave-like theory with the knowledge of the initial data on $C$. It is also
$S^{\prime \prime}=S=C^{\prime \prime}$, thus $S$ is causally closed.
 It was
shown in \cite{ceggg} that the causal closure of an achronal set includes its
 domain of dependence, and in a globally hyperbolic space-time the domain 
of dependence and the causal closure  
coincide for achronal 
sets bounded in time. 
Therefore, eq.(\ref{c}) applied to a set $S$ covering a piece of a
space-like surface $C$ implies that the algebra corresponding to $S$ includes
the algebra corresponding to the domain of dependence of $C$, as expected from
a theory with hyperbolic equations of motion. However, eq. (\ref{c}) is in
general stronger than what can be induced by causal propagation (see Section
III below).  

We will postpone to Section
III and IV the discussion of the relation between causally closed sets and
the light-sheets generated by spatial codimension 2 surfaces. 
In this work the main discussion is centered in the geometry of the nets of
causal diamond shaped sets, while its relation with the local algebras and
the counting of degrees of freedom will be heuristic. We argue that
under some physical conditions the net of sets to which are assigned the
local operator algebras have to be taken non orthomodular. We show that this
simple geometry can be associated with a possible origin of the holographic
property in an effective quantum field theory description.  We construct a
net of causal sets that implements the cutoff in a covariant way and lead to
an explanation of the non positive expansion light-sheet selection in the
covariant entropy bound.

\section{The lattices of causally closed sets}

In this Section we introduce several lattices of causal space-time subsets and
briefly investigate their properties. A more extensive analysis
for the orthomodular case  can be consulted in \cite{hc,cj}. Given a
 globally hyperbolic space-time ${\cal M}$ and a set $S$ in ${\cal M}$
we call its causal opposite or orthocomplement $ S^{\prime }$ to the set where
the local operators in a given quantum theory are constrained to commute with
the local operators in $S$. We explore several possibilities for the operation 
of causal opposite and its consequences for the lattices of the "diamond
shaped" (causally closed) sets. These sets are the ones that
satisfy $S=S^{\prime \prime}$, and thus its definition is tied to the opposite
operation.

We start with a definition for $S^{\prime }$
as the set of all points $ x$ such that there is no time-like curve connecting
$x$ with a point in $S$. Thus,
\begin{equation}
S^{\prime }=-\left( S\cup I^{+}(S)\cup I^{-}(S)\right) \,,  \label{op}
\end{equation}
where $I^{+}(S)$ and $I^{-}(S)$ are the chronological future and past of $S$,
that is, the set of points that can be reached by future directed (past
directed) time-like curves starting at a point in $S$. 
We call ${\cal L}_{T}({\cal M})$ to the set of causally closed sets
$S$,  $S^{\prime\prime }=S$, where the opposite operation is given in
equation (\ref{op}).
 
Let us look more closely to the properties of ${\cal L}_{T}({\cal M})$. The
empty set $\emptyset $ and ${\cal M}$ belong to ${\cal L}_{T}({\cal M})$ and
are mutually complementary. It is easy to show that it is
$S^{\prime \prime \prime }=S^{\prime }$ and then $S^{\prime }$ is an element 
of ${\cal L}_{T}( {\cal M})$ for any set $S$.
 The operation of taking the causal opposite
is internal in ${\cal L}_{T}( {\cal M})$. We also have 
the order relation given by the set inclusion $\subseteq $. We can
define two additional binary internal operations in ${\cal L}_{T}({\cal M})$,
the meet $\wedge $ and the join $\vee $, given by 
\begin{eqnarray} A\wedge B
&=&A\cap B\,, \label{meet}\\
 A\vee B &=&\left( A\cup B\right) ^{\prime \prime }\,. \label{join}
\end{eqnarray}

The set ${\cal L}_{T}({\cal M})$ with the order relation $\subseteq $ and
the operations $^{\prime }$, $\wedge $ and $\vee $, forms what is called an
orthocomplemented lattice \cite{hc} (see \cite{bo,ql} for the mathematical
context). A lattice ${\cal L}$ is an ordered set under 
some order relation $\subseteq$, where the greatest lower bound and the 
lowest upper bound of two elements $A$ and $B$ with respect to this order 
relation always exist, and are 
represented by the operations $A\wedge B$ and $A \vee B$ respectively. An 
orthocomplemented lattice has in addition a unary operation $A\rightarrow 
A^{\prime}$ called opposite or orthocomplement, such that
\begin{eqnarray} 
A &\wedge &A^{\prime }=\emptyset \,, \\
A &\vee & A^{\prime }=I \,, \\
A &=& A^{\prime \prime } \,, \\
A &\subseteq &B\Rightarrow B^{\prime }\subseteq A^{\prime}\,, 
\end{eqnarray}
where $I$ is maximal element of the lattice (it is ${\cal M}$ for 
${\cal L}_{T}( {\cal M})$).  

As a more familiar example of an
orthocomplemented lattice we have the set ${\cal B}(U)$ of all subsets of a
given set $U$.  There, the order relation is again given by the inclusion 
$\subseteq $, while the other  
operations are the set complement $-$, intersection $\cap $, and union 
$\cup $, respectively. The properties of the operations in
${\cal L}_{T}({\cal M})$ that come from the orthocomplemented lattice
structure copy those of ${\cal B}(U)$. For example, the operations $\wedge $
and $\vee $ are associative and symmetric. The
duality relations also hold in any orthocomplemented lattice,
\begin{eqnarray}
(A\vee B)^{\prime } &=&A^{\prime }\wedge B^{\prime }\,, \\
(A\wedge B)^{\prime } &=&A^{\prime }\vee B^{\prime }\,.
\end{eqnarray}

However, while the operations $\cap $, and $\cup $ are distributive in $
{\cal B}(U)$, the operations $\wedge $ and $\vee $ are not distributive in $
{\cal L}_{T}({\cal M})$. A similar situation occurs in the set of all closed
vector subspaces of a Hilbert space, ${\cal C}({\cal H})$. The set ${\cal C}
({\cal H})$ forms an orthocomplemented lattice under the order given
by $\subseteq $, the opposite of a subspace $V$ given by the orthogonal
space $V^{\perp }$, and the meet and join given by the intersection $\cap $,
and sum $\oplus $ of vector spaces respectively. The sum and
intersection do not distribute as can be checked with the vector spaces
generated by three independent vectors. However, a weaker form of
distributivity holds in ${\cal C}({\cal H})$ called orthomodularity, that is
a central property in the studies of quantum logic \cite{ql}.

\begin{figure}
\centering
\leavevmode
\epsfxsize=13.5cm
\epsfbox{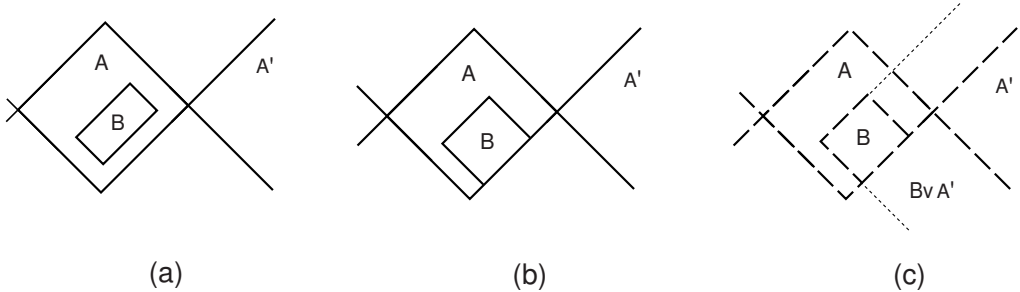}
\bigskip
\caption{The orthomodular law is valid in ${\cal L}_{T}({\cal M})$: (a)
example where $B$ is included in the interior of $A$, and (b) where $B$ is
included in $A$ and share part of the border with $A$.
 In both cases it is $A^{\prime } \vee B=A^{\prime } \cup B$ (assuming that
 $A$ and $B$ do not share part of their spatial corner), and then
$A\cap (B\vee A^{\prime })=B$. All sets include their time-like border. (c)
The orthomodular law is not valid in ${\cal L}_{\cal O}({\cal M})$. All sets
are open. The dotted line shows the set $B\vee A^{\prime }$, and it is $B
\subset A\cap (B\vee A^{\prime })$. The case where $B$ is separated from the
borders of $A$ is similar to (a), and orthomodularity holds for this special
configurations.} \end{figure}

It was shown in \cite{cj} for Minkowski space, and for general
space-times (without imposing any causality condition) in \cite{hc} that 
the set ${\cal L}_{T}({\cal M})$ with the order relation $\subseteq$ and the
operations $-$, $\vee$, and $\wedge$ defined above is an orthocomplemented
orthomodular (and non distributive) lattice  (see Fig. 2(a-b)). The
orthomodularity condition is that for any $A$ and $B$ in the lattice
\begin{equation}
 B\subseteq A\Rightarrow \,A\wedge (B\vee A^{\prime })=B\,.
\label{cuah} 
\end{equation} 

In quantum mechanics the lattice
of closed vector subspaces of the Hilbert space has a logical interpretation
in terms of physical propositions, and the orthomodular property is inherent
to this logic structure \cite{ql}. The proposition corresponding to a
subspace $V$ on a vector state $\Psi$ is given by "the orthogonal projection
of $\Psi$  onto $V$ is
$\Psi$". Thus the answer is yes if the state vector belongs to $V$ and no if
$\Psi$ belongs to the orthogonal space to $V$. The lattice of causally 
closed sets ${\cal L}_{T}({\cal M})$ have a logical interpretation in terms 
of
physical propositions given by space-time subsets \cite{hc}. The proposition 
corresponding to a
set $S$ about a point like particle is given by "the particle passes through
$S$", while the negation of that proposition is given by "the particle passes
through $S^{\prime}$". The prescription $S=S^{\prime \prime}$ is related with
the logical axiom that states that the opposite of the opposite of a
proposition gives again the same proposition. It is necessary to restrict to
the causally closed sets in order to have a logic in terms of space-time
sets, since for different types of sets the opposite of the corresponding
proposition is a statement about particle trajectories that can not be put
as a subset proposition of the above kind.

Before showing the meaning of orthomodularity in the present context we will
construct a different orthocomplemented lattice of causal sets that does not
have this property.

In the literature the operator algebras are usually associated with open
sets, because they are thought as coming from smoothed quantum fields.
It has also been suggested that the corresponding net of subsets of 
space-time is orthomodular \cite{he}. The lattice 
${\cal L}_{T}({\cal M})$ is orthomodular,
but it is not formed by open sets. We will see that one can not have both
things together.

There is another unsatisfactory feature of ${\cal L}_{T}({\cal M})$ in
this context, the fact that there are orthogonal
sets that can be joined by a null geodesic (see Fig. 2(b)). Here we use
the term orthogonal borrowed from the lattice of closed subspaces of the
Hilbert space ${\cal C}({\cal H})$ when referring to two sets $X$ , $Y$, such
that $X$ $\subseteq Y^{\prime }$. Thus, as information can be passed from one
set to the other the operators based on them should not necessarily commute.
To take into account the propagation of massless fields one would then replace
the definition of the opposite (\ref{op}) by 
\begin{equation}
S^{\prime }=-\left( J^{+}(S)\cup J^{-}(S)\right) \,, \label{equu}
\end{equation}
where $J^{+}(S)$ and $J^{-}(S)$ are the causal future and past of $S$, that
is, the set of points that can be reached from $S$ by future
directed (past directed) time-like or null curves. Again an
orthocomplemented lattice ${\cal L}_{C}({\cal M})$ is obtained picking up the
causally complete sets $S=S^{\prime \prime }$, where the opposite operation
is given by equation (\ref{equu}), and the meet and join by eqs. (\ref{meet})
and (\ref{join}) \cite{hc}. Now the light rays coming from $S$ do not
intersect $S^{\prime }$. However not all sets in ${\cal L}_{C}({\cal M})$
are open. 

We can construct another lattice of causally closed sets where opposite sets
are not causally connected, but now formed by open sets, using the opposite
given by 
\begin{equation} 
S^{\prime }=-(\overline{J^{+}(S)\cup J^{-}(S)})\,,
\label{hs} 
\end{equation}
and where $\bar{A}$ means the topological closure of $A$. 
We will call
${\cal L}_{{\cal O}}({\cal M})$ to the lattice of open sets that are causally
closed with respect to the opposite operation (\ref{hs}). As was shown in
\cite{hc},  ${\cal
L}_{{\cal O}}({\cal  M})$ is also an orthocomplemented lattice. Both lattices
${\cal L}_{{\cal O}}({\cal M})$ and ${\cal L}_{C}({\cal M})$ behave very
similarly, except that ${\cal L}_{ {\cal O}}({\cal M})$ is formed by open
sets and do not contain lower dimensional objects. Furthermore, in a
discretized version of space-time they would coincide as explained below. 

However, the lattices ${\cal L}_{
{\cal O}}({\cal M})$ and ${\cal L}_{C}({\cal M})$, in contrast to ${\cal
L}_{T}({\cal M})$, are not orthomodular. This is shown in the example of
figure 2(c). As explained in \cite{hc} the orthocomplemented structure follows
naturally from the existence of an orthogonality relation between points in
space-time, but the strongest requirement of orthomodularity seems to require
the exact definition (\ref{op}) for the opposite.

\subsection{Some consequences of orthomodularity}
There are two equivalent conditions to orthomodularity for an arbitrary
orthocomplemented lattice ${\cal L}$ that we will now illustrate. The first
is a kind of good property under reduction to a subspace. The lattice ${\cal
L}$ is orthomodular if and only if for any $A\in {\cal L}$ the family
${\cal L} _{A}$, formed by the sets $B\in {\cal L}$, $B\subseteq A$, is again
an orthocomplemented lattice,
where the opposite operation restricted to ${\cal L}_{A}$ is $
B\mid _{A}^{\prime }=B^{\prime }\cap A$. Then it follows that the lattice 
${\cal L}_{A}$ 
is also orthomodular. That this condition is true for ${\cal L}_{T}({\cal M
})$ and false for ${\cal L}_{{\cal O}}({\cal M})$ is exemplified in Fig. 3.
What this is telling is the following. An element $A$ in ${\cal L}_{T}
( {\cal M})$ is not a submanifold of ${\cal M}$ because it is not an open set,
but the elements of ${\cal L}_{T}({\cal M})$ inside $A$ form a good
orthomodular lattice with respect to the induced operations in $A$. On the
contrary, an element $A$ in ${\cal L}_{{\cal O}}({\cal M})$ is a submanifold
of ${\cal M}$ because it is open, but the elements of ${\cal L}_{{\cal O}}(
{\cal M})$ inside $A$ are not in general elements of the lattice induced by
the
reduced notion of causality ${\cal L}_{{\cal O}}(A)$. The sets in ${\cal L}_{
{\cal O}}({\cal M})$ included in $A$ that do not belong to ${\cal L}_{{\cal
O}}(A)$ share a piece of its null border with the null border of $A$, that
is, they are in contact with $A$ from inside. In fact, a weaker form of
orthomodularity for $
{\cal L}_{{\cal O}}({\cal M})$ holds, that reads
\begin{equation}
\overline{B}\subseteq A\Rightarrow \,A\wedge (B\vee A^{\perp })=B\, ,
\end{equation}
where $A$ and $B$ belong to ${\cal L}_{{\cal O}}({\cal M})$. 
The condition $\overline{
B}\subseteq A$ is sufficient for $B\in {\cal L}_{{\cal O}}(A)$,
but is not necessary since the examples suggest that whenever $B$ share a
piece of the spatial corner with $A$ it is $B\in {\cal L}_{{\cal O}}(A)$.

\begin{figure}
\centering
\leavevmode
\epsfysize=4cm
\epsfbox{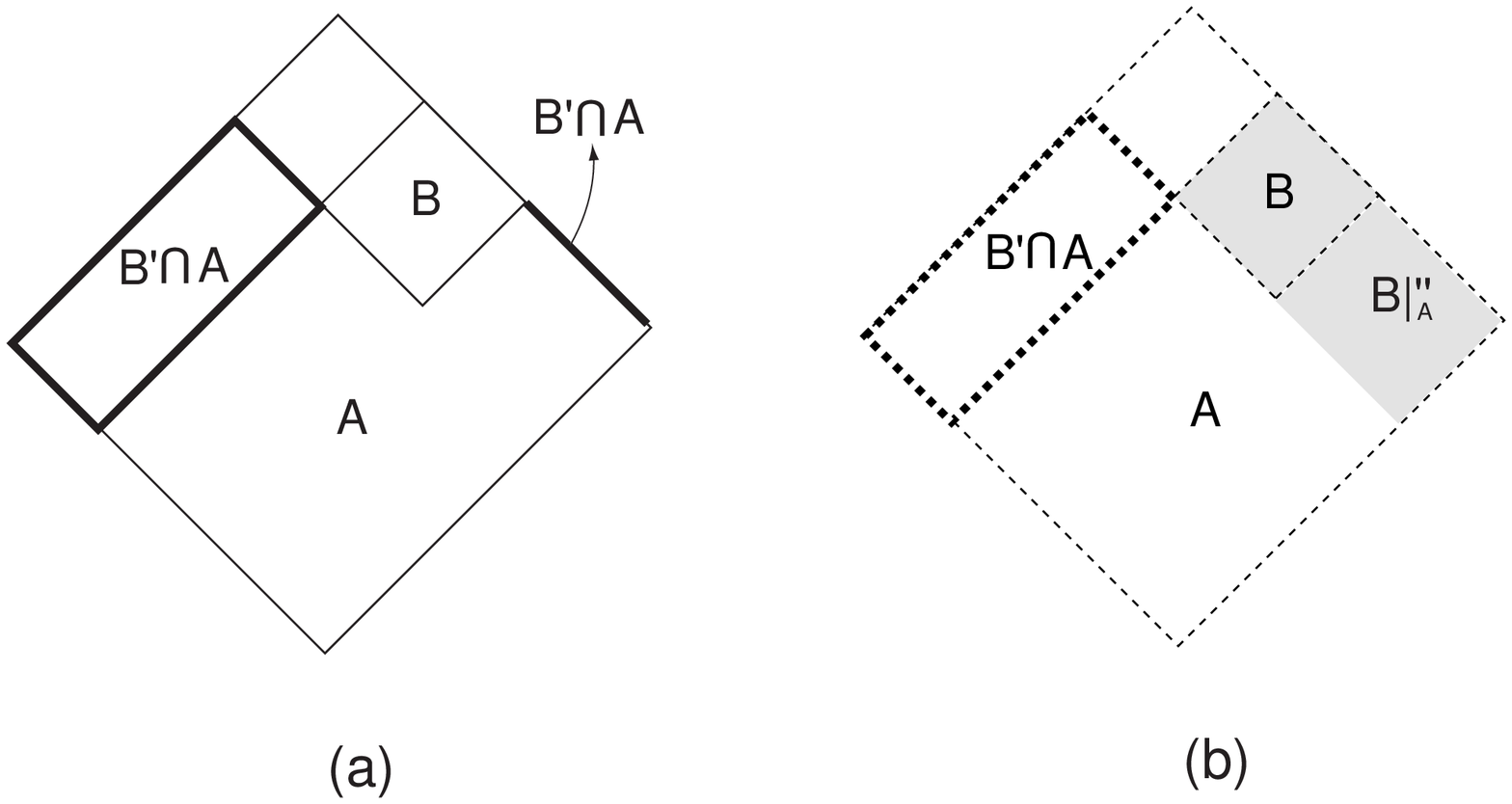}
\bigskip
\caption{(a) The family of subsets $B\subseteq A$, where $B$ and $A$ belong
to ${\cal L}_{T}({\cal M})$, form an orthomodular lattice where the opposite
is given by $B\mid_A^{\prime}=B^{\prime }\cap A$. The figure shows the
fundamental importance of the existence of null surfaces in ${\cal
L}_{T}({\cal M})$ for orthomodularity. (b) With the inherited notion of the
opposite for a set $A$ in ${\cal L}_{\cal O} ({\cal M})$ the set
$B\mid_{A}^{\prime \prime }$ (in gray) is different from $B$. Therefore
$B\in {\cal L}_{\cal O}({\cal M})$ is included in $A$ but it does not belong
to ${\cal L}_{\cal O}({\cal A})$. }
\end{figure}

This has a resemblance with what happens in the exterior of a black hole or
in the Rindler wedge in Minkowski space where the observers can restrict 
themselves to
living in a submanifold, but at the cost of dealing with a non pure state.
Here, as we reduce to a submanifold $A$ there are sets $B$ of ${\cal M}$
included in $A$ that would have associated an algebra of local operators $
{\cal A}(B)$. However, ${\cal A}(B)$ does not exist in the theory restricted
to $A$ for some of the $B\subseteq A$. Thus, some local operators in $A$
would be missing from the set of operators available to the observer.
Somehow the lattice ${\cal L}_{{\cal O}}({\cal M})$ would have information
about the entanglement of fluctuations in the vacuum across causal horizons
as implied by the Reeh-Schlieder theorem.

The second condition equivalent to orthomodularity is that for any set $S$
in an orthocomplemented lattice ${\cal L}$ there must not exist a set $T$
in ${\cal L}$ with $T\subseteq S$, $T\neq S$, and such that $T\vee S^{\prime
}={\cal M}$. That is, the complement of $S^{\prime }$ is unique among the
sets included in $S$. This condition is respected in
${\cal L}_{T}({\cal M})$ while it does not hold in ${\cal L}_{{\cal O}
}({\cal M})$. More generally,
given two orthogonal sets $A$ and $B$, and $C\subset B$, $C\neq B$, in an
orthomodular lattice it can not be that $B\vee A=C\vee A$. However, this is
true for certain sets in ${\cal L}_{\cal O} ({\cal M})$, what could have very
interesting consequences as we will see in the next Section (see Fig. 5(b)).

As the difference between the lattices presented so far is somewhat a subtlety
related to the borders of the sets, one could wonder if a slight modification
of the definitions would not yield a lattice of open sets orthomodular. This
is not possible if one wants to retain the algebraic properties of the
lattice and where the opposite consists of  points spatially
separated. For example a tentative possibility would be to identify the
operations in ${\cal L}_{{\cal O}}({\cal M})$ with operations in ${\cal L}_{
T}({\cal M})$ assigning to the set $S$ in ${\cal L}_{{\cal O}}({\cal
M})$ the set $\bar{S}$ in ${\cal L}_{T}({\cal M})$. However this
fails because a piece of a null surface is an element in ${\cal L}_{T
}({\cal M})$ without interior, and its existence is crucial for
orthomodularity (see Fig. 2(b)). Adding null surfaces to ${\cal L}_{{\cal O}
}({\cal M})$ will lead to ${\cal L}_{T}({\cal M})$, that has
 orthogonal elements causally connected, or to ${\cal L}_{C}({\cal
 M})$, that is again non orthomodular.

\subsection{Models of causal lattices for a space-time with a smallest scale}
A more convincing argument in favor of a non orthomodular
 lattice comes in a context where there is a cutoff.  For example, take a
discretized version of space-time choosing points at random with constant
medium density with respect to the volume form, and dropping the rest of the
space-time \cite{blms}. There the lattices we introduced will basically
coincide, because the differences appear only when there are points lying in
the same null geodesic, an event of zero probability. Call the resulting
lattice ${\cal L}_{D}({\cal M})$. Also, even if it happens that two points 
are
light connected, in any discretized version, the fact that their operators
should not commute, and the points should not be orthogonal becomes 
strengthened in comparison with the continuum. Thus, it seems that when
space-time itself is blurred at some scale (but retaining a causal relation
as in \cite{blms}), extremely localized objects as the border of sets would
have no meaning. The causal structure would survive at large scales through a
 non orthomodular causal net.  We will not study here the interesting problem
of finding the lattice operations of this type of discretized space-time in
the thermodynamical limit. We only note that for simple examples of
${\cal L}_{D}({\cal M})$ constructed with a few points, the result is
somewhat intermediate between the behavior of ${\cal L}_{\cal O}({\cal M})$
and ${\cal L}_{T}({\cal M})$ (see figure 4), a feature that we will encounter
again in Section IV for a different type of regularized lattice that we now
introduce.

\begin{figure}
\centering
\leavevmode
\epsfysize=5cm
\bigskip
\epsfbox{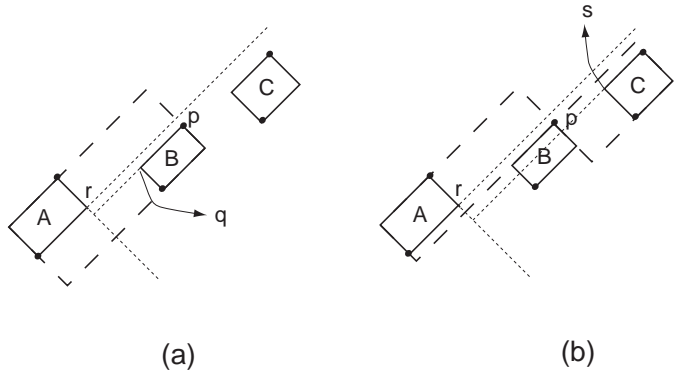}
\caption{The lattice formed by choosing points at random with constant
medium density in space-time and retaining the causal relations. The sets
formed with the points that are in the intersection of the past and the
future of two time-like
separated points, as the sets $A$, $B$ and $C$, are always in the lattice.
(a) There are no points inside the set bounded by the light rays passing
through
$q$ and $r$. Thus, the join of $A$ and $B$ is as in the lattice of open sets.
(b) There is the point $p$ inside the set bounded by the light rays passing
through $s$ and $r$. The set $A\vee C$ results cut by the past and future
of $p$ as compared with what happens in case (a).}
\end{figure}

We mentioned that for a theory of point particles, or more generally any
 theory that allows to probe the space-time structure up to the continuum,
the lattice ${\cal L}_{T}({\cal M})$ has an interpretation in terms of
physical propositions given by space-time subsets. However, if there is a
lowest scale that can be tested then the situation is different. Suppose that
we can probe the space-time up to a minimal size
$\mu$. We can think that our probe is an extended particle of rest frame 
size $\mu$. 
Then, given a set $S$ we can probe that a point is spatially separated
 from $S$ only if it is separated from $S$ by a spatial distance greater 
than $\mu$. Equivalently, two points will be causally unrelated if any of the
 extended particle trajectories passing over one point does not reach the 
other. Thus we define the opposite operation by 
\begin{equation} 
S^{\prime
}=\{ x/ x\neq y \ $and$ \ d^2(x,y)\leq -\mu^{2} \ $for all$ \ y\in S \} \ ,
\label{muop}
 \end{equation}
where we use the signature $(1,-1,-1,-1)$ for the metric and 
$d^{2}(x,y)$ is the greatest square geodesic distance between $x$ and $y$
(in case $x$ and $y$ can not be joined by any geodesic set
$d(x,y)=-\infty$). The use of $\leq$ or $<$ in (\ref{muop}) is not relevant
in what follows. A structure of orthocomplemented lattice is obtained for
the causally closed sets from this definition for the opposite and we call the
resulting lattice   ${\cal L}_{-\mu^{2}}({\cal M})$.

We have that
${\cal L}_{-\mu^{2}}({\cal M})$
is a non orthomodular lattice, while the lattices
 produced by choosing a positive value instead of $-\mu^{2}$ in (\ref{muop})
behave as orthomodular for simple examples. The non orthomodularity
 of ${\cal L}_{-\mu^{2}}({\cal M})$  is in accord with the known 
result that given a orthocomplemented structure the orthomodularity is 
related with the presence of a set of states that is enough to separate the 
different propositions \cite{ql}. 

With the interpretation of the opposite of a set $S$ as the set of points
where the operators commute with those based on $S$ we arrive at the same
 operation (\ref{muop}) in the case where there is a lowest fundamental
scale. Operators algebras based on regions separated by a spatial distance
smaller than $\mu$ may not commute because that would imply that a smaller
scale have an observational significance.
Only a finite number of mutually orthogonal sets
are included in a given bounded diamond in the lattice
${\cal L}_{-\mu^{2}}({\cal M})$,
so the net can be thought as a covariant way of implementing a cutoff because 
independent degrees
of freedom based on different space-time sets should have commuting
generators.

The causal closure (the double opposite) in ${\cal L}_{T}({\cal M})$
for an achronal bounded set 
has also the interpretation of being its causal domain of
dependence. A similar  interpretation has the closure in ${\cal
L}_{-\mu^{2}}({\cal M})$. Given an orthogonal set of points $C$ in this
lattice we can draw a Cauchy surface $C_{1}$ passing through all of them. Then
given a point $x$ in the generated set $C^{\prime \prime}$ and any extended
particle trajectory that passes over $x$, the trajectories cut $C_{1}$ at
points which have a spatial distance from $C$ smaller that $\mu$. Thus, $C$
acts as a generalized form  of Cauchy surface for $C^{\prime \prime}$.   

Thus, it seems that the presence of a cutoff in
the space-time description leads to a non orthomodular lattice for the
causally closed sets that implement the causal law (\ref{c}). Here the 
causal closure has to be calculated with the opposite (\ref{muop}).
  
From their definitions we see that
${\cal L}_{T}({\cal M})$, ${\cal L}_{C}( {\cal M})$, and ${\cal L}_{{\cal
O}}({\cal M})$ do not change with conformal transformations, and thus, they
are a property of the conformal structure.  On the contrary, ${\cal
L}_{D}({\cal M})$ and ${\cal L}_{-\mu^{2}}({\cal
M})$ are not conformally invariant.

\section{Non orthomodularity and the covariant entropy bound}

We will assume that we can construct the algebra ${\cal A}(S)$
corresponding to a set $S$ formed by set union of orthogonal sets $S_{i}$ 
with
the elements of the corresponding algebras. Thus
\begin{equation}
{\cal A}(\cup _{i}S_{i})=\vee _{i}{\cal A}(S_{i})\ ,  \label{d}
\end{equation}
where the sets $S_{i}$ are orthogonal, and ${\cal A}_{1}\vee {\cal A}_{2}$
denotes the algebra generated by ${\cal A}_{1}$ and ${\cal A}_{2}$. 
This is a natural postulate that involves a local principle. The operators 
associated to a set of space-like separated regions can be constructed with
operators based on the given regions. However, Eq.(\ref{d})
can have problems in  theories with
global non gauge charges (see \cite{he}, specially Section III.4). 
Remarkably,
these charges are not supposed to survive at a fundamental level or when all
effective terms in the Lagrangian are taken into account.

From eqs. (\ref{c}) and (\ref{d}) it follows that
\begin{equation}
{\cal A}(\vee_i S_{i})=\vee_i {\cal A}(S_{i}),  \label{law}
\end{equation}
for orthogonal families $S_i$ of elements. 

Here we note that we used eq. (\ref{c})
only for the union of orthogonal families of causally closed sets.
Otherwise this equation can not be justified on the basis of causal
propagation only. For example the double opposite of a set $S$ formed by two
time-like displaced diamonds is a set bigger than the domain of dependence of
$S$, that is, the set of all points through which every inextendible
time-like curve intersects $S$. 

 Equation (\ref{law})
is  the statement that the local operators acting in a region of space-time
can be constructed from the mutually commuting sets of operators in regions
non causally connected to each other. The content of this equation clearly 
depends on the lattice chosen for the causally closed sets. It expresses in
algebraic manner the causality of dynamical evolution due to classical 
space-time geometry when using the lattice
${\cal L}_{T}({\cal M})$, and it is a natural generalization 
for the case where the causal structure is only described by a lattice. We
will proceed assuming that (\ref{law}) is valid and show that it has 
strong implications for a non orthomodular lattice.  

Now we see the role of the
lattice of causally closed sets that is used as the base for the theory. In
the case of ${\cal L}_{T}({\cal M})$, the picture is that of hyperbolic
dynamical laws with the initial data set on Cauchy surfaces. In fact, given a
diamond $S$ in ${\cal L}_{T}({\cal M})$ if we try to form it as a join of
mutually orthogonal smaller diamonds, we find  that these later always cover
a Cauchy surface for $S$ (see Fig. 5(a)). Thus ${\cal A} (S)$ can be seen as
formed by mutually commuting generators on the Cauchy surface.

The lattice ${\cal L}_{{\cal O}}({\cal M})$
implies a very different counting of degrees of freedom. As can be seen in
Fig. 5(b),
there each diamond is generated by a subset arbitrarily close to the
spatial border of $S$, plus an arbitrarily small orthogonal diamond near one
of the tips. Thus, most of the independent degrees of freedom are localized
in the surface $\Omega$. When there is a cutoff the number of degrees of
freedom would increase as the bounding area. The role of the non orthomodular
behavior is clear. In figure 5(b) it is $A\vee C=A\vee B=S$, where $C$ is
a proper subset of $B$. As already mentioned this situation can not happen in
an orthomodular lattice.

\begin{figure}[t]
\centering
\leavevmode
\epsfysize=5cm
\epsfbox{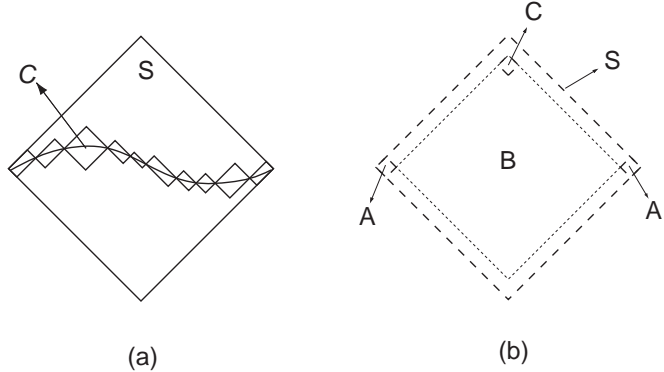}
\bigskip
\caption{ Forming $S$ as the join of mutually orthogonal sets. (a) In ${\cal
L}_{T}({\cal M})$ the mutually orthogonal sets have to cover a Cauchy surface
$C$ for $S$. (b) In ${\cal
L}_{\cal O}({\cal M})$ it suffices to take a small set $A$ covering the
spatial corner of $S$ and one orthogonal set $C$ near the future or past tip.
 Note that $A\vee B=A\vee C=S$.}
\end{figure}

A similar construction may be implemented for a null hypersurface
converging to a point $q$, and orthogonal to a non necessarily closed
codimension 2 spatial surface $\Omega$. There, a neighborhood of the null
hypersurface can be formed as the join of a small set near $\Omega$ and a
small orthogonal diamond near $q$.

Thus, it seems that a non orthomodular causal net captures the features
of the holographic property. However, the lattice ${\cal L}_{\cal
O}({\cal M})$ has two drawbacks. One is that it does not tell how to count
degrees of freedom, as an arbitrarily small set in the lattice includes
infinitely many orthogonal subsets. The other is that it is conformally
invariant. Then, the lattice makes no difference between expanding and
contracting null hypersurfaces orthogonal to $\Omega$ if they finally
converge to a point. So, the same argument that applies for the surface $H$
in the case of Fig. 1(a) would apply for the dashed line representing a null
hypersurface in Fig. 1(b), leading to entropy bounds with simple 
counterexamples (for example if the universe if big enough beyond the Hubble 
radius as suggested by inflation).

These difficulties can be cured using ${\cal
L}_{-\mu^2}({\cal M})$ that implements a finite cutoff. We can give a 
measure of a set $S$ in this
lattice that would correspond to the number of degrees of freedom
 available inside $S$. First, we want the independent degrees of freedom to
be assigned to orthogonal sets. Second, if a set contains a pair of
orthogonal subsets it can not be assigned just one degree of freedom. Thus we
can think in the sets that do not contain any pair of orthogonal subsets as 
the
building blocks. For definiteness we will take the points, that do not
have any proper subsets, as such building blocks. These are the atoms of the 
lattice. The minimal
number $N(S)$ of orthogonal points we need to generate $S$ (or a set that
covers $S$ in general) is then naturally associated with some constant times
 the number of degrees of freedom in $S$. This definition is covariant.

As an example consider three dimensional Minkowski space, and the
diamond set $S$ in Fig. 1(a). It is possible to construct the net of
orthogonal points that cover the Cauchy surface $C$ as in Fig. 6(a).
It also generates its domain of dependence $S$. The number of points then
grows with the volume of the Cauchy surface. On the other hand, as shown in
Fig. 6(b), with points near the bounding area $\Omega$ plus one single
orthogonal point one can also generate a set covering $S$. The number of
points is smaller than in the previous case, thus it represents the
actual number (or a greater bound) of degrees of freedom.

\begin{figure}[t]
\centering
\leavevmode
\epsfysize=5cm
\epsfbox{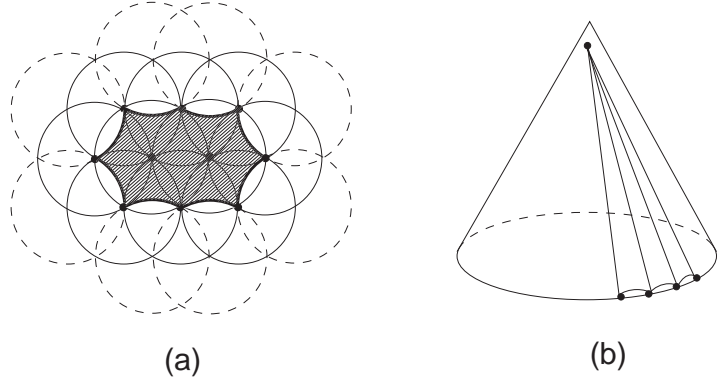}
\bigskip
\caption{(a) Generating a spatial flat surface with a set of orthogonal
points in ${\cal L }^{+}_{-\mu^2 }({\cal M })$. The circles are of radius
$\mu$. Thus the opposite of the set of points is outside the solid circles.
 The dashed circles are centered in the points of the opposite best
positioned to take out most of the set when doing the double orthogonal. The
shaded area is the set generated (or the double orthogonal) by the points,
shown in the plane of the points. Here we have set the distance between
neighbor points as the minimum to be orthogonal. In fact, the most efficient
 manner to cover the two dimensional surface is to take a hexagonal net like
 the shown in the picture, but where the distance between neighbors is
infinitesimally near and smaller than
$\sqrt{3} \mu$.
When covering a
surface in this way and including the time this set of points will also
generate a set approximating the causal development of the surface. (b) When
the radius of a circle is much bigger than $\mu$ the set formed by the
central point and points in the circle will not generate a surface because
some orthogonal points can be situated inside the circle. However, we can
lift the central point of the circle in time, reducing its distance to the
circle to the minimum $\mu$. Then, if the distance between neighbor points in
the circle is less than $2\,\mu$, they will generate a set approaching a
null surface.  When the circle is covered, and only then, they will generate
the whole cone and the lower cone also (not shown).}
\end{figure}

The reason for the holographic reduction of the number of independent 
degrees of freedom
 can be seen more easily looking at Fig.(7). Suppose we have a covariant way
 of assigning degrees of freedom to the Cauchy surfaces $C_{1}$ and 
$C_{2}$, for example by separating them by some spatial distance as 
we have done using the lattice ${\cal L}_{-\mu^{2}}({\cal M})$. Then, moving
 $C_{2}$ to approach the null boundary of $S$ its volume goes to zero
 reducing the number of independent degrees of freedom that the surface can 
hold. As both Cauchy surfaces must have the same number of independent 
degrees of freedom since they describe the same physics, most of the 
degrees of freedom of $C_{1}$ must not be independent. However, the 
degrees of freedom in the spatial corner may well turn out to be independent 
when restricting attention to the particular diamond set $S$.

\begin{figure}[t]
\centering
\leavevmode
\epsfysize=5cm
\epsfbox{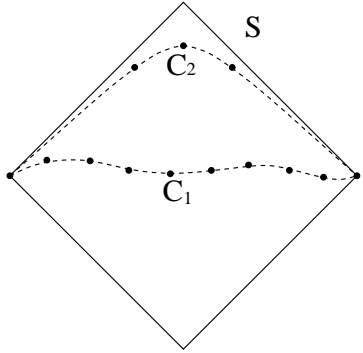}
\bigskip
\caption{Two Cauchy surfaces $C_{1}$ and $C_{2}$ for the 
diamond shaped set $S$. The points represent discrete degrees of freedom 
equally spaced on the Cauchy surfaces. As $C_{2}$ approaches the null 
border of $S$ its volume tends to zero and the number of points that it can 
hold is reduced.}
\end{figure}

In the following Section we will give evidence in the context of a simple
example that the lattice ${\cal L}_{-\mu^2}({\cal M})$ does indeed choose the
non positive expansion condition for the applicability of the entropy bound.

\section{The lattice ${\cal L}_{-\mu^2}({\cal M})$ for three dimensional 
Robertson-Walker models}

In this Section we consider spatially flat three dimensional
Friedman-Robertson-Walker models with a metric given by
\begin{equation}
ds^{2}=\eta ^{\beta }\left( d\eta ^{2}-dr^{2}\right) \,, \label{dsd}
\end{equation}
where $\eta $ is the conformal time, $dr^{2}=dx^{2}+dy^{2}$, and the
exponent $\beta $ is taken in the range $0\leq \beta <\infty $. We focus on 
 spatial one dimensional surfaces $\Omega$ formed by an arc of a circle at
fixed $\eta$. We show that the set generated doing the double orthogonal in 
the lattice  ${\cal L}_{-\mu^2}({\cal M})$ of the set formed by  
$A(\Omega)/(2\mu)$
orthogonal points along $\Omega$ plus a single additional orthogonal point,  
approximates the light 
sheets of negative expansion corresponding
to $\Omega$ in the limit of small
$\mu$. Here $A(\Omega)$ is the area of $\Omega$. On the contrary, for
positively expanding null congruences orthogonal to $\Omega$ the 
implementation of this construction requires a greater number of points by 
unit surface than in the non expanding case, or it simply can not be realized 
with orthogonal points.

This metric (\ref{dsd})
can be expressed in the form
\begin{equation}
ds^{2}=dt^{2}-a(t)^{2}dr^{2},
\end{equation}
and the relation between time and conformal time is given by
\begin{eqnarray}
t &=&\frac{\eta ^{1+\frac{\beta }{2}}}{^{1+\frac{\beta }{2}}}\,, \\
a(t) &=&\left( \frac{1}{2}(2+\beta )t\right) ^{\frac{\beta }{\beta +2}}\,.
\end{eqnarray}
{}From here we see that the Hubble radius $a/\dot{a}$ is given by
$R_{H}=\frac{ 2 \eta^{(1+\beta / 2)}}{\beta }$. The case $\beta =0$ corresponds
to flat space-time. 
The geodesic equation for a curve $\eta (r)$ is
\begin{equation}
\eta ^{^{\prime \prime }}-\frac{\beta }{2}\eta ^{-1}\left( \eta ^{\prime
\,2}-1\right) =0\text{.}
\end{equation}
The solution for null geodesics is simply
\begin{equation}
\eta =\pm r+C\,,
\end{equation}
where $C$ is a constant. We will be interested in small deviations from the
null geodesics with $\eta$ growing with $r$, then we write
\begin{equation}
\eta (r)=r+C+\delta (r)\,,
\end{equation}
with $\delta (r)\ll 1$. The linearized geodesic equation is
\begin{equation}
\delta ^{^{\prime \prime }}-\frac{\beta }{r+C}\,\delta ^{\prime }=0\text{.}
\end{equation}
The solution of this equation that departs from the point $r=0$, $\eta =C$ is
\begin{equation}
\eta =r+C+\frac{C_{1}}{\beta +1}\left( \left( r+C\right) ^{(\beta
+1)}-C^{(\beta +1)}\right) \,,  \label{equ}
\end{equation}
where $C_{1}$ is a small constant, negative for space-like geodesics and
positive for time-like geodesics. The square distance along these geodesics
from $r=0$ to $r=r_{0}$ is
\begin{equation}
s^{2}=\left( \int_{0}^{r_{0}}dr\,\eta ^{\frac{\beta }{2}}\sqrt{\eta ^{\prime
\,2}-1}\right) ^{2}\,,
\end{equation}
and its first order expression in $C_{1}$ is given by
\begin{equation}
s^{2}=\frac{2C_{1}}{(\beta +1)^{2}}\left( \left( r_{0}+C\right) ^{(\beta
+1)}-C^{(\beta +1)}\right) ^{2}\,.  \label{ese}
\end{equation}

We have now all the elements for analyzing the following geometry.
Let the spatial surface $\Omega$ be an arc of circle of radius $y_{0}$ in
the plane $\eta=1$. Let $\Phi$ be the a set of points along $\Omega$
distanced $2 \varepsilon$ between neighbors, where $\varepsilon$ is a small
quantity of the order of the cutoff scale $\mu$, and covering in this way
$\Omega$. We are assuming that the size of $\Omega$ is much greater than the
cutoff $\mu$.

Take two
neighbor points $p_{\pm }$ in $\Phi$, and without loss of generality assume
that they have coordinates $p_{\pm }=\left( \pm \varepsilon ,0,1\right) $
(see Fig. 8). The intersecting null geodesics coming
from $ p_{+}$ and $p_{-}$ will intersect each other at the plane $x=0$. Any
pair of these intersecting null geodesic with intersection point $q$ will
form the edges of a null surface $\omega $ converging to $q$ and
bounded below by the sector of $\Omega$ between $ p_{+ }$ and $p_{-}$. Let 
the
point $q$ be given by the coordinates $\left( 0,\,y_{0},\,1+y_{0}+\varepsilon
^{2}/(2y_{0})\right) $ written at second order in $\varepsilon $. Thus,
$q$ is at null distance from all points in $\Phi$. The
transversal area on the null surface $\omega $ at the coordinate $y$ is $
2\varepsilon \frac{(y_{0}-y)}{y_{0}}(1+y)^{\beta /2}$. This is increasing for
$y=0$ when $y_{0}>2/\beta =R_{H}$. In such case the null surface $\omega$ is
not a light sheet in the sense of \cite{bousso}.

\begin{figure}
\centering
\leavevmode
\epsfysize=5cm
\epsfbox{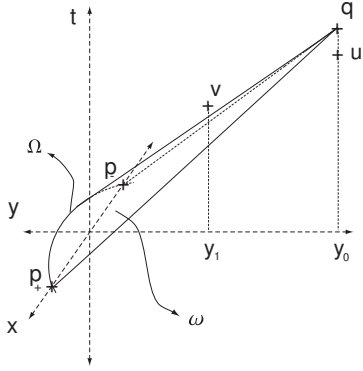}
\bigskip
\caption{The geometry used in this Section. The points $p_{+}$ and $p_{-}$ 
are situated along a line parallel to the $x$ axis at $t=1$ with coordinate
$x=\pm \epsilon$. Two light rays coming from $p_{\pm}$ intersect at $q$ in
the plane
$x=0$ at coordinate $y=y_0$. The point $u$ is at coordinate $y=y_0$ and at
distance square $-\mu^2$ from $p_{\pm}$. The point $v$ is in the plane $x=0$,
at coordinate $y=y_1$, and very near the null surface $\omega$.}
\end{figure}

We intend to approximate $\omega $ using the set generated by $\Phi$
 and an additional point $u$
orthogonal to $\Phi$ and very near $q$. The best chance is taking $u$ to
have the
maximal distance square required to be orthogonal to $ p_{\pm }$,
$d^{2}(u,p_{\pm })=-\mu ^{2}$. Using (\ref{ese}) with $C=1$ and $
r_{0}=y_{0}$ to zero order ($C_{1}$ is already small), we obtain $C_{1}$, and
replacing in (\ref{equ}) we have \begin{equation} u=\left(
0\,\,\,\,\,,\,\,\,\,y_{0}\,\,\,\,\,\,,\,\,\,\,\,1+y_{0}+\frac{ \varepsilon
^{2}}{2y_{0}}-\mu ^{2}\frac{(\beta +1)}{2\left( (1+y_{0})^{(\beta
+1)}-1\right) }\right) \,.  \end{equation}

  Now we want to explore the
conditions under which $u$ and $\Phi$ generate a set approximating $\omega $.
 If we find that a point
$v$, very near the surface $\omega $, belongs to the orthogonal of $\{u,\,
p_{+},\,p_{-}\}$ (and then to the orthogonal of $u\cup \Phi$), then the
generated set will not contain a hole in $\omega $ below or above $v$,
because all past and future of $v$ is not included in $v^{\prime }$.

The form of the
generated set $ ( u \cup \Phi )^{\prime \prime}$
will be shaped as the intersection of the opposites of the points in
$( u \cup \Phi )^{\prime }$. If such a point $v$ near $\omega$ does not
 exist, then it can be seen that the points in $(u \cup \Phi)^{\prime}$
that could impede the generation of a set approximating $\omega$
must be near the extension of the null generators of $\omega$
beyond $\omega$. It is easy to see that if there are
points in $\Phi$ along a non infinitesimal arc on each side of $p_{+}$ and
$p_{-}$, there are no points in $(u\cup \Phi)^{\prime}$ near the null
generators of $\omega$ extended at the future of $\omega$.
Therefore, in the case no such a point $v$ can be found, a point near
$\omega$ will be generated for each $y$
between $0$ and $y_{0}$.

Thus, we search points $v$ capable of making holes in the generated set,
and we situate $v$ with a coordinate $y=y_{1}$ and on the plane $x=0$. The
reason for this last election is that the distance from $u$ is the same in
the whole arc of radius $ y_{0}-\,\,y_{1} $ around $y_{0}$, while if one
chooses a point $v$ with some $x\neq 0$ there is less distance square to one
of the points $p_{+}$ or $ \,p_{-}$ and more to the other. Thus, as the
distance square from $v$ to both has to be less than $-\mu ^{2}$, the best
chance is with $x=0$. Points $v$ displaced along the $x$ direction beyond the
surface $\omega$ can be orthogonal to $p_{+}$ and $p_{-}$, but they are
taken into account in the next
patch of null surface corresponding to other points in $\Phi$.

Thus we
choose
\begin{equation}
v=\left( 0\,\,\,\,\,,\,\,\,\,\,y_{1}\,\,\,\,\,,\,\,\,\,\,1+y_{1}+\frac{
\varepsilon ^{2}}{2y_{0}}+\Delta \right) \,,\,
\end{equation}
where $\Delta $ is a small quantity and $0<y_{1}<y_{0}$. Using (\ref{equ})
and (\ref{ese}) we obtain the distance square $d^{2}(v,p_{\pm })$ that is
given by
\begin{equation}
d^{2}(v,p_{\pm })=-\frac{2}{(\beta +1)}\left( (1+y_{1})^{(\beta
+1)}-1\right) \left( \varepsilon ^{2}\frac{(y_{0}-y_{1})}{2y_{1}y_{0}}
-\Delta \right) \,.
\end{equation}
Likewise, we compute the distance square between $v$ and $u$,
\begin{equation}
d^{2}(v,u)=-\frac{2}{(\beta +1)}\left( (1+y_{0})^{(\beta
+1)}-(1+y_{1})^{(\beta +1)}\right) \left( \frac{\mu ^{2}(\beta +1)}{2\left(
(1+y_{0})^{(\beta +1)}-1\right) }+\Delta \right) \text{.}
\end{equation}
As mentioned, for making holes in the generated set we have to
demand $d^{2}(v,p_{\pm })\leq -\mu ^{2}$ and $d^{2}(v,u)\leq -\mu
^{2}$. These translates into the following conditions for $\Delta $
\begin{eqnarray}
&&\varepsilon ^{2}\frac{(y_{0}-y_{1})}{2y_{1}y_{0}}-\frac{\mu ^{2}(\beta +1)
}{2\left( (1+y_{1})^{(\beta +1)}-1\right) }\geq \Delta \,,  \label{aaa} \\
\Delta &\geq &\mu ^{2}\frac{(\beta +1)}{2}\left[ \frac{1}{\left(
(1+y_{0})^{(\beta +1)}-(1+y_{1})^{(\beta +1)}\right) }-\frac{1}{\left(
(1+y_{0})^{(\beta +1)}-1\right) }\right] \,.
\end{eqnarray}
Given $\beta $ and $y_{0}$ we have to satisfy these conditions for $y1\in
[0\,,\,y_{0}]$. Thus, $\Delta \ge 0$. If a point $v$ is in
$\{u,$ $p_{+},\,p_{-}\}^{\prime }$, then all the points
in the null surface $\omega $ for $x=0$ and smaller $y$ will be time like
connected with $v$ and will not be in the generated set.

The condition for the existence of $\Delta$ can be restated as
\begin{eqnarray}
&\,&\frac{1}{\left( (1+y_{0})^{(\beta +1)}-1\right) }-\frac{1}{\left(
(1+y_{1})^{(\beta +1)}-1\right) }-\frac{1}{\left( (1+y_{0})^{(\beta
+1)}-(1+y_{1})^{(\beta +1)}\right) } \nonumber \\
&\,&+\frac{\varepsilon ^{2}}{\mu ^{2}}\frac{
(y_{0}-y_{1})}{(\beta +1)\,y_{1}y_{0}}\geq 0\,. \label{func}
\end{eqnarray}

For a given $\beta >0$, taking $y_{0}\rightarrow \infty $, $y_{1}\rightarrow
\infty $, and $y_{0}/y_{1}\rightarrow \infty $, we see that there are
always solutions for any $\varepsilon $ when the surface is big enough. Thus,
surfaces with big $y_{0}$ are not generated given a fixed $\varepsilon$.
 For $\varepsilon >
\mu$ we have points in $(u\cup \Phi)^{\prime}$ in the initial surface and a
 set converging to $\omega$ for small $\mu$ will not be generated. Let
us take the minimal set of orthogonal elements in the spatial surface that
do not admit the addition of other orthogonal points, that is, we take
$\varepsilon$ to be an infinitesimal smaller than $\mu$. With such
a choice the function (\ref{func}) is decreasing with $y_{1}$. Then it
suffice to take the limit $y_{1}\rightarrow 0$. There the condition
(\ref{func}) becomes
\begin{equation}
 y_{0}\geq \frac{2}{\beta }\text{. }
\end{equation}

Therefore, for expanding surfaces, the number of orthogonal
points required in the spatial surface $\Omega$ to generate the approximate
null
surface will be greater than the minimal orthogonal set of points that cover
(but not generate by themselves) the surface $\Omega$ \cite{rech}. This does
not happen for contracting surfaces, where it suffices with taking
$A(\Omega)/(2\mu)$ points on the surface $\Omega$ of area $A(\Omega)$.

For $\mu / 2 \le \varepsilon < \mu$ we can still generate the set converging
to $\omega$ in certain range of $y_{0}$, with the point $u$ plus orthogonal
points in the initial surface, leading to an
extension of the holographic idea, but where the number of degrees of freedom
per unit area is increased with respect to the standard value applicable to
the non expanding case. For a large enough $y_{0}$ it is not possible to
choose $\varepsilon$ in this range to generate the surface. Of course it can
still be generated with non orthogonal points, for small enough
$\varepsilon$. Numerically, certain conservation of the number of points 
seems
to hold. For a given $y_{0}$ let $\hat{\varepsilon}$ be the maximal
$\varepsilon$ such that there exist no $y_{1}$ such that Eq.(\ref{func})
holds. Then let the area of maximal expansion for the initial surface of area
$2\, \hat{\varepsilon}$ be $A_{\max}(2\, \hat{\varepsilon})$. It is
$A_{\max}(2\, \hat{\varepsilon})/(2\, \mu)\simeq 1$. Therefore the number of
points needed in the surface $\Omega$ (in general non orthogonal points) is
very similar to the minimal number of orthogonal points needed to cover the
surface of maximal expansion.

We hope that an analysis on the same line involving the geodesic deviation
equation would yield the result of this Section on the relation between the
covariant entropy bound and the geometry given by the lattice 
${\cal L}_{-\mu^{2}}({\cal M})$,
 in a
more general context. 
\section{Conclusions}

We have seen, considering the geometry of classical space-time, that
there exist two well differentiated classes of
orthocomplemented nets of causal sets. One is given by the orthomodular net
${\cal L}_{T}({\cal M})$, which is related to the usual picture of using
independent data on Cauchy surfaces. The other class is of non orthomodular
nets, that can be related to a reduction of degrees of
freedom of the holographic type.  We have argued that the second type of 
nets 
should be used for constructing algebraic quantum theories if there is a 
cutoff scale. 

Somewhat at the extreme of the non orthomodular behavior is the lattice
${\cal L}_{\cal O} ({\cal M})$, that, being conformally invariant, does not
differentiate between contracting and expanding light-sheets. When
regularizing the lattices one takes into account the metric in addition to
the causal structure, and the resulting behavior is intermediate between
 ${\cal L}_{T}({\cal M})$ and ${\cal L}_{\cal O}({\cal M})$.

We have
constructed the non orthomodular lattice ${\cal L}_{-\mu^2}({\cal M})$
that implements a covariant cutoff for the causal lattices. This allows
a geometrical definition of the number of degrees of freedom for an arbitrary
 set in space-time. We have seen that this number is consistent with the 
Bousso
covariant entropy bound in a simple example, where it reproduces the non
negative expansion condition for the election of light-sheets.

An important point suggested by the lattices constructed in this work is that
while the covariant entropy bound would hold in its original form, the
holographic projection for a diamond shaped set or a light sheet would not be
simply to degrees of freedom on the spatial bounding surface. These
 are most of the independent degrees of freedom, but something
more seems to be needed along the light sheet or its tip to close the
algebra.

The lattice ${\cal L}_{T}({\cal M})$ can be taken as a description of the 
causal structure in classical space-time \cite{hc}. It is othomodular and it 
also has an interpretation in terms of physical propositions for classical 
theories. As we have argued the corresponding object that describes causality
 when space-time can not be taken classical at all the scales should be a 
non orthomodular lattice. If the causal propositions remain  
physical propositions that would mean that orthomodularity is lost as a 
property of the quantum logic structure. A change of the quantum mechanics
 postulates in a cosmological scenario is advocated for example in 
Ref.\cite{qml}.  

It would be interesting to explore the purely geometrical problem posed by the
lattice ${\cal L}_{-\mu^2}({\cal M})$. Does it reproduces the holographic
 property for a general space-time?. 
A precise formulation for this statement is the
following. Given a codimension 2 spatial surface $\Omega$ and one of its
light-sheets $H$, and let $N(S)$ be the minimal number of orthogonal points
needed to generate a set covering $S$ in space-time. Then we would like to
test if
$\lim_{\mu \rightarrow 0} \frac{N(\Omega)}{N(H)}=1$, or find the
conditions for its validity.

More generally, the lattice ${\cal L}_{-\mu^2}({\cal M})$ suggests
a generalized geometrical version of
entropy bound. This is simply that the entropy in a set $S$ has to be less
than a constant times $N(S)$. The constant has to be adjusted to match the
covariant bound when appropriate, and is given by
$3\sqrt{3}\mu^2 /(8G)$ in four dimensions \cite{rech}. Thus, this geometrical
bound would be independent of $\mu$ when the cutoff is taken smaller than all
the curvature scales in the set.

In relation to this idea arises the question of what kind of regularization
could lead to the stronger form of
covariant entropy bound given in \cite{fmw}, which is related to the 
Bekenstein
entropy bound \cite{bek}, and implies the generalized second law. This
 can not be 
deduced from the counting of degrees of freedom by the lattice
${\cal L}_{-\mu^2}({\cal M})$.

 It would also be interesting to
investigate other regularizations, as the given by the lattice resulting from
a random distribution of points in space-time ${\cal L}_{D}({\cal M})$ in the
thermodynamical limit.

 Here we have assumed a base classical space-time and a net of local
 algebras of operators in Hilbert space as a generalized form of quantum
field theory. This structure should appear above some distance scale. What
seems to be odd is that, following what we have argued, the covariant entropy
bound would hold in a form logically independent of the Einstein equations
for the metric. After all the Einstein equations are essential for curving
the space in such a way to save the bound in several examples 
\cite{bousso,fmw}. 
However, the
order of the implications could possibly be inverted using an idea by
Jacobson \cite{jac}. 
In that work the Einstein equations are deduced starting from the area law 
for the
entropy, and using the second law of thermodynamics as seen by accelerated 
observers to relate a heat flux given by the stress 
tensor with the area expansion. Thus, it seems likely that to 
implement an effective theory valid above some smallest fundamental scale 
gravity should appear as a requirement of self consistency.

\acknowledgments

I acknowledge very useful discussions with C. Rovelli and correspondence with
 R. Bousso. This work was supported by CONICET, Argentina.

\end{document}